\documentstyle[12pt,epsfig]{article}

\catcode`\@=11

\def\bold#1{\setbox0=\hbox{$#1$}%
\kern-.025em\copy0\kern-\wd0
\kern.05em\copy0\kern-\wd0
\kern-.025em\raise.0433em\box0 }

\def\lsim{\mathrel{\mathpalette\@versim<}}
\def\gsim{\mathrel{\mathpalette\@versim>}}
\def\@versim#1#2{\vcenter{\offinterlineskip
\ialign{$\m@th#1\hfil##\hfil$\crcr#2\crcr\sim\crcr } }}

\newcount\@tempcntc
\def\@citex[#1]#2{\if@filesw\immediate\write\@auxout{\string\citation{#2}}\fi
  \@tempcnta\z@\@tempcntb\m@ne\def\@citea{}\@cite{\@for\@citeb:=#2\do
    {\@ifundefined
       {b@\@citeb}{\@citeo\@tempcntb\m@ne\@citea\def\@citea{,}{\bf ?}\@warning
       {Citation `\@citeb' on page \thepage \space undefined}}%
    {\setbox\z@\hbox{\global\@tempcntc0\csname b@\@citeb\endcsname\relax}%
     \ifnum\@tempcntc=\z@ \@citeo\@tempcntb\m@ne
       \@citea\def\@citea{,}\hbox{\csname b@\@citeb\endcsname}%
     \else
      \advance\@tempcntb\@ne
      \ifnum\@tempcntb=\@tempcntc
      \else\advance\@tempcntb\m@ne\@citeo
      \@tempcnta\@tempcntc\@tempcntb\@tempcntc\fi\fi}}\@citeo}{#1}}
\def\@citeo{\ifnum\@tempcnta>\@tempcntb\else\@citea\def\@citea{,}%
  \ifnum\@tempcnta=\@tempcntb\the\@tempcnta\else
   {\advance\@tempcnta\@ne\ifnum\@tempcnta=\@tempcntb \else \def\@citea{--}\fi
    \advance\@tempcnta\m@ne\the\@tempcnta\@citea\the\@tempcntb}\fi\fi}
\catcode`\@=12 % at signs are no longer letters
\def\naive{na\"{\i}ve}

\def\nextline{\hfill\break}
\def\refnextline{\nextline} % so that can go back to no line breaks

\def\half{\hbox{${1\over2}$}}
\renewcommand{\bar}[1]{\overline{#1}}

\newcommand{\ket}[1]{\,\left|\,{#1}\right\rangle} 
\newcommand{\etal}{{\em et al.}}
\newcommand{\ie}{{\it i.e.}}

\def\ubar{\bar{u}}
\def\dbar{\bar{d}}
\def\sbar{\bar{s}}
\def\cbar{\bar{c}}
\def\qbar{\bar{q}}
\def\Qbar{\bar{Q}}
\def\pbarp{\bar{p}\hskip0.02em p}
\def\rhopi{\rho\hskip0.08em\pi}

\begin{document}
\baselineskip=16pt
%\begin{flushleft}
%\draft
%\end{flushleft}
\begin{flushright} SLAC--PUB--7463\\ TAUP 2417-97\\ hep-ph/9704379\\ April 1997
\end{flushright}
\bigskip\bigskip

\thispagestyle{empty}
\flushbottom

\centerline{\Large\bf Intrinsic Charm of Vector Mesons:} \vspace{5pt}
\centerline{{\Large\bf A Possible Solution of the
``\mbox{\boldmath $\rhopi$} Puzzle''}
\footnote{\baselineskip=14pt  Work supported in part by the Department
of Energy, contract DE--AC03--76SF00515.}}

\vspace{22pt}
\centerline{\bf Stanley J. Brodsky}
\vspace{8pt}
\centerline{\it Stanford Linear Accelerator Center}
\centerline{\it Stanford
University, Stanford, California 94309}
\centerline{\it e-mail:
sjbth@slac.stanford.edu} \vspace{8pt}
\centerline{and}
\vspace{8pt}
\centerline{\bf Marek Karliner}
\vspace{8pt}
\centerline{\it School of Physics and Astronomy} \centerline{\it Raymond and
Beverly Sackler Faculty of Exact Sciences} \centerline{\it Tel-Aviv University,
69978 Tel-Aviv, Israel} \centerline{\it e-mail: marek@vm.tau.ac.il}
\vspace*{1cm}

\centerline{ABSTRACT}

An outstanding mystery of charmonium physics is why the $J/\psi$ decays
prominently to pseudoscalar plus vector meson channels, such as $J/\psi
\rightarrow \rhopi$ and $J/\psi \rightarrow K^*K$, whereas the $\psi^\prime(2S)$
does not. We show that such decays of $J/\psi$ and their suppression for
$\psi^\prime(2S)$ follow naturally from the existence of intrinsic charm
$\ket{\qbar q \, \cbar c}$ Fock components of the light vector mesons.

\vfill
\centerline{Submitted to Physical Review Letters.}
\vfill
\newpage
%--------------------- end of title page, text begins here ------------

\baselineskip=19pt

One of the basic tenets of quantum chromodynamics is that heavy quarko\-nium
states such as the $J/\psi, \psi^\prime,$ and $\Upsilon$ must decay into light
hadrons via the annihilation of the heavy quark constituents into gluons, as
shown in Fig.~1(a). This assumption is motivated by the OZI rule which
postulates suppression of transitions between hadrons without valence quarks in
common. A central feature of the PQCD analysis is the fact that the annihilation
amplitude for quarkonium decay into gluons occurs at relatively short distances
$r \simeq 1/m_Q$, thus allowing a perturbative expansion in a small QCD coupling
$\alpha_s(m_Q)$.

In this letter we shall challenge the assumption that quarkonium states
necessarily decay via intermediate gluon states. We shall argue, in analogy with
the analysis\cite{EGK,EKKS} of the nucleon strangeness content, that the
wavefunctions of the light hadrons, particularly vector mesons such as the
$\rho$ and $K^*$, have a non-negligible probability to have higher Fock state
components containing heavy quark pairs \cite{IC}. The presence of intrinsic
charm and bottom in the light hadron wavefunctions then allows transitions
between heavy quarkonium and light hadrons by rearrangement of the underlying
quark lines, rather than by annihilation.

One of the most dramatic problems confronting the standard picture of quarkonium
decays is the $J/\psi \to \rhopi$ puzzle \cite{puzzle}. This decay occurs with a
branching ratio of $(1.28\pm 0.10)\%$ \cite{RPP96}, and it is the largest
two-body hadronic branching ratio of the $J/\psi$. The $J/\psi$ is assumed to be
a $\cbar c$ bound state pair in the $\Psi(1S)$ state. One then expects the
$\psi^\prime = \Psi(2S)$ to decay to $\rhopi$ with a comparable branching ratio,
scaled by a factor $\sim 0.15$, due to the ratio of the $\Psi(2S)$ to
$\Psi(1S)$ wavefunctions squared at the origin. In fact, $B(\psi^\prime
\to \rhopi) < 3.6\times 10^{-5}$ \cite{BES96}, more than a factor of 50
below the expected rate. Most of the branching ratios for exclusive
hadronic channels allowed in both $J/\psi$ and $\psi^\prime$ decays
indeed scale with their lepton pair branching ratios, as would be
expected from decay amplitudes controlled by the quarkonium wavefunction
near the origin,
\begin{equation}
{B(\psi^\prime \to h)\over B(J/\psi \to h)}\simeq 
{B(\psi^\prime \to e^+e^-)\over B(J/\psi \to e^+e^-)} 
=0.147\pm0.023 \quad\cite{RPP96,BES96}
\label{Bscaling}
\end{equation}
where $h$ denotes a given hadronic channel.
The $J/\psi\to \rhopi$ and $J/\psi \to K K^*$ decays also conflict dramatically
with PQCD hadron helicity conservation: all such pseudoscalar/vector two-body
hadronic final states are forbidden at leading twist if helicity is conserved at
each vertex \cite {BL,BLT}.

The OZI rule states that hadronic amplitudes with disconnected quark lines are
suppressed; in QCD this corresponds to the assumption that there is a numerical
suppression of amplitudes in which multiple-gluon intermediate states occur.
Although the OZI rule has provided a useful guide to the general pattern of
hadronic reactions involving strange particle production, there are glaring
exceptions: for example, experiments at LEAR have found \cite{LEARexps} that in
the $\pbarp$ annihilation at rest the OZI-violating ratio 
$B(\pbarp \to \phi \pi/
\pbarp \to \omega \pi)$ is enhanced by almost two orders of magnitude compared
to the \naive\ OZI expectations, and that the process $\pbarp \to \phi \phi$
occurs at roughly the same rate as $\pbarp \to \omega \omega$.

The absence of OZI suppression can be understood \cite{EGK,EKKS} if one takes
into account the presence of {\em intrinsic strangeness} in the proton, \ie, one
allows for $\ket{u u d \,\sbar s }$ Fock components in the proton
wavefunction.\footnote{The Fock state expansion may be rigorously defined in a
frame-independent way using light-cone Hamiltonian methods
\protect\cite{BLreview}.} The intrinsic strange quarks are part of the hadronic
composition of the proton in distinction to {\em extrinsic strangeness} arising
from simple gluon splitting. The $\pbarp \to \phi \pi$ and $\pbarp \to \phi
\phi$ amplitudes can then occur simply by rearrangement diagrams in which the
strange quarks initially present in the incoming $p$ and $\bar p$ appear as the
valence components of
$\phi$ mesons in the final state. The OZI rule is {\em evaded}, since the
annihilation of the $p $ and $\bar p$ into intermediate gluons is in fact not
required.

It is clearly interesting to extend these considerations to the charm and bottom
sector. In general, the probability to find heavy quarks or high mass
fluctuations in the light hadron wavefunctions which are multi-connected to the
valence constituents is suppressed by inverse powers of the relevant mass. For
example one can use PQCD to show that the probability for intrinsic charm or
bottom Fock states $\ket{u u d \,\Qbar Q}$ in the proton wavefunction scales as
$1/m^2_Q$ \cite{BH}. The light cone wavefunctions for such states, $\Psi^p_{uud
\,\Qbar Q}(x_i, k_{\perp i}, \lambda_i)$, peak at the smallest invariant mass of
the partons; \ie, at equal rapidity for the constituents. Thus the heavy quarks
tend to have the largest momentum fractions $x_i = k^+_i/p^+ \propto
m_{\perp i } = (m^2_i + k^2_{\perp i })^{\half}.$ In fact the EMC
experiment which measured the charm structure function of the nucleon
found an excess of events at large $Q^2$ and $x_{bj}$ well beyond what
is expected from photon gluon fusion.  Analysis shows that the EMC data
are consistent with an intrinsic charm probability of $0.6 \pm 0.3\%$
\cite{HarrisVogtSmith}.
There is also a recent interesting
proposal to apply these ideas in order to 
reconcile the recent HERA data with the Standard Model \cite{ICatHERA}.

An interesting test of intrinsic charm in the proton would be a search for
$\pbarp \to J/\psi\, J/\psi$,
$\pbarp \to \phi\, J/\psi$, $\pbarp \to \omega J/\psi$ above the charm
threshold, processes which can occur without annihilation into gluons and thus
without OZI suppression because of the presence of charm and strangeness in the
initial state. Similarly, exclusive open charm reactions such as 
$\pbarp \to \bar\Lambda_c \hskip0.02em\Lambda_c $ can occur through 
rearrangement of the initial charm quark lines.

The discussion and the experimental evidence for the intrinsic charm is usually
phrased in terms of the charm content of the nucleon. On the other hand, there
is a well known and highly successful phenomenological constituent quark model
in which the nucleon contains just three constituent quarks. In order to
reconcile the two physical pictures, one is inevitably led to the conclusion
that the constituent quarks are themselves complicated composite objects,
containing a sea of gluons, light $\qbar q$ pairs and a small, but
non-negligible $\cbar c$ intrinsic charm component. In addition, intrinsic
contributions are produced from diagrams which are multiconnected to two or more
valence quarks in the nucleon. This then immediately implies that the vector
mesons, such as $\rho$, $K^*$, etc., also contain an intrinsic charm component,
for they are built from the same constituent quarks as the baryons.

The presence of intrinsic charm in light hadrons can also have important
consequences \cite{EFHK} for the exclusive hadronic decays of $D$- and
$B$-mesons, which are usually analyzed by assuming only valence quarks
in hadronic states. Any hadron containing a light quark would also be
expected to have higher Fock states containing heavy quark pairs by the
same type of quantum fluctuations which produce intrinsic strangeness
and charm in the nucleon. The surprisingly large branching ratios $D \to
\phi K$ is possibly due to this effect \cite{EFHK}.

Let us now re-examine the $J/\psi \to \rhopi$ decay, allowing for intrinsic
charm in the wavefunctions of the final state hadron. For example, consider the
light-cone Fock representation of the $\rho$: $\rho^+ = \Psi^\rho_{u \dbar}
\,\ket{u \dbar} \,+\, \Psi^\rho_{u\dbar\,\cbar c}\,\ket{u \dbar\,\cbar
c}\,+\,\cdots\ \ .$ The
$\Psi^\rho_{u \dbar \cbar c} $ wavefunction will be maximized at minimal
invariant mass; \ie\ at equal rapidity for the constituents and in the spin
configuration where the $u \dbar$ are in a pseudoscalar state, thus minimizing
the QCD spin-spin interaction. The $\cbar c$ in the $\ket{u \dbar\,\cbar c}$
Fock state carries the spin projection of the $\rho.$ We also expect the
wavefunction of the $\cbar c$ quarks to be in an $S$-wave configuration with no
nodes in its radial dependence, in order to minimize the kinetic energy of the
charm quarks and thus also minimize the total invariant mass.

The presence of the $\ket{u \dbar\,\cbar c}$ Fock state in the $\rho$ will allow
the $J/\psi \to \rhopi$ decay to occur simply through rearrangement of the
incoming and outgoing quark lines; in fact, the $\ket{u \dbar \,\cbar c}$ Fock
state wavefunction has a good overlap with the radial and spin $\ket{\cbar c}$
and $\ket{u \dbar}$ wavefunctions of the $J/\psi$ and pion. 
Moreover, there is no conflict 
with hadron helicity conservation, since the $\cbar c$ pair in the
$\rho$ is in the $1^-$ state. 
On the other hand, the overlap with the
$\psi^\prime$ will be suppressed, since the radial wavefunction of the $n = 2$
quarkonium state is orthogonal to 
the node-less $\cbar c$ in the $\ket{u \dbar \,\cbar c}$ 
state of the $\rho$. This simple argument provides a 
compelling explanation of the absence of $\psi^\prime \to \rhopi$ and other 
vector pseudoscalar-scalar states.\footnote{The possibility that the radial
configurations of the initial and final states could be playing a role in the
$J/\psi \to \rhopi$ puzzle was first suggested by S. Pinsky \cite{Pinsky}, who
however had in mind the radial wavefunctions of the light quarks in the $\rho$,
rather than the wavefunction of the $\cbar c$ intrinsic charm components of the
final state mesons.}

We can attempt to make a rough
estimate of the decay rate $J/\psi \to \rho\pi$ by comparing it with the
measured rate of the analogous decay $\phi \to \rhopi$,
$\Gamma(\phi \to \rhopi) \approx 6\times 10^{-4}$ GeV \cite{RPP96}, assuming
that the latter also occurs via coupling to the intrinsic $\sbar s$ component in
the $\rho$.
Consider the Feynman graph where an  $Q\bar Q$ is connected
to two valence quarks in the wavefunction of the hadron through two hard gluons.
This gives a factor of $\alpha^2_s(M_Q^2)$ in the amplitude and thus
$\alpha^4_s(M_Q^2)$ in the probability.  The same factor occurs in the
rearrangement decay rate shown in figure 1(b), since there is implicitly a hard
gluon connecting the $c$ with the $u$ and the $\bar c$ with the $\bar d$ in the
$\rho$ wavefunction.  Thus, qualitatively, we can estimate that the ratio of
probabilities for intrinsic charm to intrinsic  strangeness in a light hadron is
of order
\begin{equation}
R_{({c \bar c / s \bar s})}
\simeq {m^2_s\over m^2_c} {\alpha^4_s( M_c^2) \over \alpha^4_s(M_s^2)},
\label{IQestimate}
\end{equation}
which is of the order of $10^{-3}$. This is also consistent in order of
magnitude with the estimates of the ratio of intrinsic charm to strangeness
obtained from deep inelastic scattering on the nucleon.
The actual numerical value is uncertain due to the uncertainties in the values
of the mass parameters and the running of the coupling at low scales. There may
be other suppression factors from the evolution of the light hadron
wavefunctions, higher order corrections, etc. In the case of in scattering
reactions with probes of low resolution, there is an additional screening of the
intrinsic sea\cite{chev,BH},  but this type of suppression does not apply to
decay amplitudes computed from the overlap of wavefunctions.

The ratio  of decay rates for $J/\psi \to \rho\pi$ to  $\phi \to
\rhopi$ from quark rearrangement should roughly scale with $R_{({c \bar c / s
\bar s})}$ times phase space, assuming that the
integration over the quarkonium wavefunctions
give similar probabilities.\footnote{In the case of the 
intrinsic charm or intrinsic strangeness
rearrangement contribution,  we only need to
compute the overlap of the light-cone wavefunctions. 
Thus there is no extra $\rho$
form factor suppression beyond the penalty to find intrinsic charm
with large invariant
mass of order of the $J/\psi$ mass in the $\rho$ wavefunction.}

This rough estimates implies $\Gamma(J/\psi \to \rho\pi)\sim 10^{-6}$ GeV,
which is consistent with the measured rate of $10^{-6}$ GeV.

Our analysis utilizes the fact that quantum fluctuations in a QCD bound state
wavefunction will inevitably produce Fock states containing heavy quark pairs.
The heavy quark pairs arising from perturbative gluon splitting are the
extrinsic contributions associated with the substructure of the gluons; the
probability for such pairs depends logarithmically on the ultraviolet resolution
scale. In the case of charmonium decay to light hadrons, the extrinsic heavy
quark fluctuations provide hard radiative corrections to the usual $c \bar c$
annihilation amplitude.

On the other hand, the intrinsic heavy quarks arise from quantum
fluctuations which are multiconnected to the valence quarks of the light
hadrons, and the wavefunctions describing these configurations will have
maximal amplitude at minimal off-shellness and minimal invariant mass.
In the case of the $\rho$ meson the $\ket{\dbar u\, \cbar c}$
wavefunction will thus be maximized when the configuration of the quarks
resembles that of a $\ket{\pi\,J/\psi}$ intermediate state, rather than
a higher mass $\ket{\bar D\,D}$ state. This preference for the lowest
invariant mass induces a relatively strong coupling
$g_{J/\psi\hskip0.1em\rho \pi}$\ , {\em i.e.} there is a natural overlap between
a $\rhopi$ and $J/\psi$ which facilitates the $J/\psi \to \rhopi$ decay, as
schematically illustrated in Fig.~1(b). The decay of the $\psi^\prime$
is naturally suppressed due to the node in its radial wavefunction, also
shown schematically in Fig.~1(c). Similarly, the $\ket{\ubar s\, \cbar
c}$ Fock component of the $K^*$ will have a favored $J/\psi\,K$
configuration, allowing the $J/\psi \to K^* K$ decay to also occur by
quark line rearrangement, rather than $\cbar c$ annihilation.

Intrinsic charm in the pion will also allow the decay $J/\psi(1S) \to
\rhopi$ to proceed through quark rearrangement diagrams. In this case the decay
can utilize configurations of the pion's $\ket{\dbar u \cbar c}$ Fock state
which resemble $\rho\,J/\psi$, where the $\rho$ and $J/\psi$ have opposite
helicity. Again, $\psi(2S) \to \rhopi$ decay will be suppressed because of the
suppressed overlap of the radial $\cbar c$ wavefunctions.

The branching ratios for the $J/\psi(1S)$ and $\psi(2S)$ for many hadronic
channels track fairly well with their leptonic branching ratios, as would be
expected if $\cbar c$ annihilation into gluons and/or photons is dominant and
unsuppressed by helicity selection rules. For example, the vector meson - scalar
meson two-body decay channels $J/\psi(1S) \to V S$ can proceed through $\cbar c$
annihilation. Note that the $\cbar c$ rearrangement contribution to $J/\psi(1S)
\to V S$ is disfavored: the $J/\psi$-scalar intrinsic charm excitation in a
vector meson wavefunction is fairly massive, and it is thus relatively
suppressed compared to the $J/\psi$-pseudoscalar excitations. On the other hand,
tensor mesons could have an appreciable intrinsic charm content. In general, a
full analysis of each exclusive decay channel will require taking into account
both $\cbar c$ annihilation and rearrangement diagrams as well as their
interference.

At first sight, the decay of $J/\psi$ to pseudovector--scalar should be
helicity suppressed in PQCD for the same reason $J/\psi$ to
pseudoscalar--vector is suppressed \cite{BL}.  The argument is that there is
only one Lorentz invariant, parity-conserving amplitude, and this requires
that the pseudovector have helicity $\pm 1$.  However, the light quark and
antiquarks emerging from the $c \bar c$ annihilation into gluons have
opposite helicity.

It is important to note that the pseudovector and scalar states are
dominantly $P$-wave bound states of light quarks.  The nonzero helicity
of the pseudovector meson can arise from the orbital angular momentum,
and thus unlike the pseudoscalar-vector channels, there is no strong
PQCD suppression of the annihilation amplitude due to helicity
conservation.  However, the form factor suppression comparing $\psi(1S)$
and $\psi(2S)$ pseudovector-scalar decays is stronger than normal
because $P$-wave wavefunctions vanish at the origin.  Thus it is possible
that both the $\cbar c$ annihilation and intrinsic charm rearrangement 
mechanisms will contribute significantly to such decay amplitudes.

It would also be interesting to compare branching ratios for the $\eta_C(1S)$
and $\eta_C(2S)$ as clues to the importance of $\eta_C(1S)$ intrinsic charm
excitations in the wavefunctions of light hadrons. In principle, similar
analyses can be carried out for exclusive $\Upsilon(1S)$ and $\Upsilon(2S)$
decays as clues to the intrinsic $b \bar b$ content of light hadrons.

Thus a systematic comparison of the various hadronic channels of heavy
quarkonium could provide important constraints on the quantum numbers,
magnitudes, and configurations of the intrinsic heavy quark excitations in light
hadron wavefunctions.

The existence of non-OZI rearrangement mechanisms for exclusive $J/\psi$ decay
will inevitably also effect the total inclusive rate for $J/\psi$ decay, and
thus modify the value of $\alpha_s$ obtained by assuming that the decay
amplitude is due solely to $\cbar c$ annihilation~\cite{alphaFromPsi}.

\bigskip
\begin{flushleft} 
{\large\bf Acknowledgments}
\end{flushleft}

The authors wish to thank CERN and the organizers of the Workshop on Strange
Structure of the Nucleon for their hospitality and for providing a stimulating
environment which catalyzed this work. This research was supported in part by
the Israel Science Foundation administered by the Israel Academy of Sciences and
Humanities, and by a Grant from the G.I.F., the German-Israeli Foundation for
Scientific Research and Development.

\def\PL{{\em Phys. Lett.\ }}
\def\NP{{\em Nucl. Phys.\ }}
\def\PR{{\em Phys. Rev.\ }}
\def\PRL{{\em Phys. Rev. Lett.\ }}

\vfill\eject
\def\mystrut{\hbox{\vrule height8.5pt depth10pt width0pt}}
\def\thefigure{1}
\begin{figure}[H]
\begin{center}
$\phantom{a}$
\vskip-1cm
\mbox{\epsfig{file=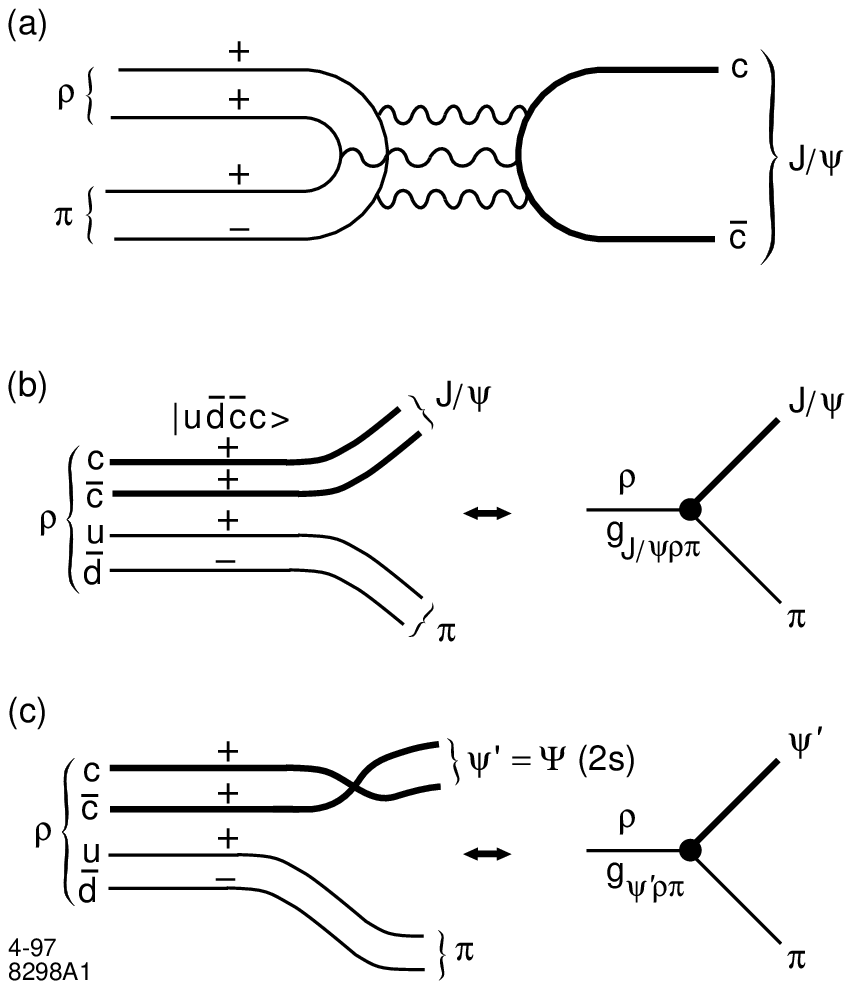,height=12truecm}}
\vskip2truecm
\caption{\protect\small
\baselineskip=15pt
\mystrut
\refnextline
(a) The decay $J/\psi (J_z = 1) \to \rhopi$ via
the standard PQCD $c \bar c$ annihilation mechanism.
A~light quark helicity-flip is required, since the $\rho$ must be produced
with helicity $\pm 1.$\mystrut
\refnextline
(b) A connected quark rearrangement diagram
which induces the $g_{J/\psi \rhopi}$ coupling, via
the higher Fock state of the $\rho$, $|u\dbar\,\cbar c\rangle$.
The $+/-$ signs on the quark lines denote the helicities
of the corresponding quarks.
In the dominant intrinsic charm Fock state of the $\rho$,
the $u\dbar$ and $\cbar c$
components of the $\rho$ are in $0^-$ and $1^-$ states, respectively,
thus generating maximal overlap with the $\pi$ and $J/\psi$ spin
wavefunctions.
\mystrut
\refnextline
(c) A ``twisted" connected diagram, schematically indicating the
suppression of $\psi^\prime \rhopi$ coupling due to the mismatch
between the nodeless wavefunction of the $\cbar c$
in the $|u\dbar\,\cbar c\rangle$ Fock state of the $\rho$ and the
one-node $2S\,$ $\cbar c$ wavefunction of the $\psi^\prime$.
}
\label{fig1}
\end{center}
\end{figure}
\end{document}